\documentclass{elsart}
\usepackage{timesfonts}
\usepackage{psfig}
\usepackage{natbib}
\def\astrobj#1{#1}
\begin{document}
\begin{frontmatter}

\title{Beam Models for Gamma-Ray Bursts Sources: Outflow Structure,
Kinematics and
Emission Mechanisms}    

\author{Enrico Ramirez-Ruiz\thanksref{email}}
\address{Institute of Astronomy, Madingley Road, Cambridge,
CB3OHA, England}
\thanks[email]{E-mail: enrico@ast.cam.ac.uk}
\and
\author{Nicole M. Lloyd-Ronning\thanksref{email1}}
\address{Canadian Institute of Theoretical Astrophysics, 60 St. George
Street, Toronto, M5S 3H8, Canada}
\thanks[email1]{E-mail: lloyd@cita.utoronto.ca}

\begin{abstract}
The variety of $\gamma$-ray burst phenomenology could be largely
attributable to differences in the opening angle of an  isotropic
outflow or to a standard type of event viewed from different
orientations. Motivated by this currently popular idea, we study the
effects of varying the energy per unit solid angle in the unsteady 
expelled outflow by an increase either in the bulk Lorentz factor or in the
baryon loading. We apply these models to 
interpret the observed correlations between variability, luminosity
and spectral peak energy and find that while the latter scenario
fails to provide a good description, bursts  produced by collisions
between similar mass shells but with increasingly large Lorentz
factors are both more variable and have larger peak spectral energies.
We present detailed internal shock calculations  confirming this
interpretation and discuss the roles various timescales, radii  and
the optical thickness of the wind play in determining this wide range
of behaviors. Finally, we discuss
the variety of scenarios in which large variations of the source
expansion velocity are naturally expected.

\end{abstract}

\end{frontmatter}

\section{Introduction}
The discovery of afterglows in recent years (Costa et al. 1997; van
Paradijs et al. 1997; Frail et al. 1997) has moved the study of
gamma-ray bursts (GRB) to a new plane. It not only has extended
observations to longer timescales and other wavelength bands, enabling the 
measurement of redshift distances, but has also
provided a direct determination of the source 
size  and confirmation of its relativistic expansion (see Piran
1999 and M\'esz\'aros 2001 for recent reviews). Detailed predictions
of the afterglow properties,
made in advance of the observations (M\'esz\'aros \& Rees 1997; Vietri
1997),  agreed well with subsequent detections at these wavelengths,
followed over timescales of months. 
While progress has been made in understanding
how the GRB and afterglow radiation arises in terms of  an unsteady
relativistic outflow followed by the development of a blast wave
moving into the external medium, interest continues to grow as new
observations provided new challenges of interpretation.

Many GRBs are now found to be associated with star forming regions (Kulkarni et
al. 1998; Fructher et al. 1999; Berger, Kulkarni \& Frail 2001), the
remnants of a massive stellar progenitor system (Piro et al. 2000; Amati et
al. 2000) and possibly supernovae (Galama et al. 1998; Bloom et
al. 1999; Reichart 1999; Bj\"ornsson et al. 2001; Lazzati et al. 2001b). 
These observations give support to the idea that the most common
GRBs could  be linked to the collapse of massive stars (Woosley 1993;
Paczy\'nski 1998; MacFadyen \& Woosley 1999). This  speculation is
made more intriguing by the recent 
report that GRBs could be highly collimated,  so that there may  be many
more events Doppler boosted in directions other than ours, bringing the
GRB rate to within a factor of 100 or so of the supernova rate and
their derived explosive bulk energies down to $\approx 10^{52.5}$ erg (Frail
et al. 2001, Panaitescu \& Kumar 2001; Piran et al. 2001; assuming a
conversion efficiency of 1\%). The
apparent dispersion in isotropic energies is then caused by a
distribution of jet opening angles -- bursts with higher energy
per solid angle are more collimated (or maybe viewed closer to the jet
axis; see Rossi, Lazzati \& Rees 2002; Salmonson \& Galama 2002; 
Zhang \& M\'esz\'aros 2002). 
New evidence and new puzzles were added when the total isotropic 
luminosity was found to be also related with a variety of temporal and
spectral behaviors exhibited by $\gamma$-ray light curves - namely the
degree of variability or ``spikiness'' (Fenimore \& Ramirez-Ruiz
2002a), the
differential time lags for the arrival of burst pulses at different
energies (Norris, Marani \& Bonnell 2000, Norris 2002), and the rest frame GRB
peak energy (Lloyd-Ronning \& Ramirez-Ruiz 2002).

These correlations are still tentative, but if confirmed they 
could be used to relate the properties of material moving
almost directly towards us (e.g. the variable activity of the $\gamma$-ray
light curve) with the ejection in directions away from our line of
sight (e.g. degree of collimation; see Salmonson 2000; Ioka \&
Nakamura 2001; Kobayashi et al. 2002; Plaga 2001). This could turn out to be 
very important, since at present we can only infer the energy per solid
angle; as yet the constraints on the angle-integrated  $\gamma$-ray
energy are not strong. The gamma-rays we receive come only from
material whose motion is directed within $1/\Gamma$ of our line of
sight. They therefore provide no information about the ejecta in other
directions. At observer times of more than a week, the beaming and
aberration effects are less extreme so we observe afterglow
emission from a wider range of angles. At
these later times, however, all memory of the initial time-structure of the
outflow responsible for the $\gamma$-ray emission would be lost.\\ 

Information regarding how much energy
and momentum has been injected, its distribution in angle, the
mass fractions in shells with different Lorentz factors and its
efficiency in radiated gamma-rays are at the forefront of attention.   
We address some aspects of these issues here by studying the  
physical conditions under which   simulated GRBs arising from internal
shocks in relativistic winds  can account for this burst
phenomenology - degree of collimation, variable $\gamma$-ray activity, the
rest frame peak energy and the total burst luminosity. 
To that effect, we consider anisotropic  
relativistic outflows where changes in the energy per solid angle
are caused either by an increase in the bulk Lorentz factor or in the
baryon loading. We apply these models to interpret the large variety
of behaviors exhibited by both the afterglow and the prompt
emission, and discuss their possible use for predicting the 
internal structure of the collimated outflow.  Some possible
scenarios in which large variations of the  energy per solid angle are
naturally expected are highlighted, along with the types of
observation that would discriminate among the various models. We assume  $H_0 =
65\,\, {\rm km} \, {\rm s}^{-1} \, {\rm Mpc}^{-1}$, $\Omega_{\rm
matter}=0.3$, and $\Omega_{\Lambda}=0.7$.

\section{Anisotropic or Highly Beamed Outflows?}

The high Lorentz factors and energies seen in GRBs are consistent
with the catastrophic formation of a stellar black hole of a few
$M_\odot$, with $\approx$ 1 \% going to a relativistic outflow.
This could be the extreme example of the asymmetric explosion
produced by supernova (Khokhlov et al. 1999), in which instead of
halting at the neutron star stage, the 
collapse continues to the black-hole stage, producing an even faster
jet in the process (MacFadyen \& Woosley 1999). GRBs arising from a
very small fraction of stars that undergo this type of  catastrophic
energy release are likely to produce collimated outflows.

Even if the outflow is not highly collimated, some beaming is
expected because energy would channeled preferentially along the
rotation axis. Also, one would expect baryon contamination to be
lowest near the axis, because angular momentum flings material away
from the axis and material with low-angular momentum falls into the
black hole. The dynamics, however, are complex. While numerical
simulations of collapse scenarios can address the fate of the bulk of
the matter (MacFadyen \& Woosley 1999; MacFadyen, Woosley \& Heger
2001; Aloy et al. 2000; Zhang et al. 2002, in preparation), higher
resolution simulations of the outer layers of the stellar mantle seem to be
required since even a very small amount of baryons polluting the
outflow could severely limit the attainable Lorentz factor.
It is quite possible, for
instance, that  the stellar pressure will tend to
collimate the fireball into  a jet. A broad spread of Lorentz factors
is thus expected -- close to the rotation axis $\Gamma$ may be 
high. At larger angles away from the jet axis, there may be an increasing
degree of entrainment, with a corresponding decrease in
$\Gamma$. So in this model, the measured flux is more intense when
observed closer to the jet axis (with an angle $\theta_v \approx 0$ from the
jet axis). A large spread in the inferred luminosities thus originates
from the differences in viewing angles (see Figure 1a). This
interpretation may be appealing because a large spread of 
both Lorentz factors and luminosities could be attributable to a
standard type of event viewed from different orientations (Rees
1999). On the other hand, the large variations in the
fluences of GRB could also originate from differences in the opening
angle $\theta_j$ of an  isotropic outflow 
(see Figure 1b; Frail et al. 2001; Panaitescu \&
Kumar 2001; Piran et al. 2001). The dynamics of the blast-wave in
the off-axis anisotropic jet are almost indistinguishable from the
uniform jet model, provided that the energy per unit
solid angle varies as $\theta_{v,j}^{\approx -2}$ (Kobayashi et
al. 2002; Rossi et al. 2002;  Salmonson \& Galama 2002; Zhang \&
M\'esz\'aros 2002).\\ 

The correlations between variability, the
luminosity per unit solid angle and the characteristic photon 
energy in the cosmological rest frame (Lloyd-Ronning \& Ramirez-Ruiz
2002) impose severe restrictions on any emission model. The
question at hand is whether this burst phenomenology,
combined with the intrinsic variations in energy per unit solid angle discussed
above, can be understood in terms of internal shocks and if so, under
what conditions. To this end, the results presented in \S 3 are
independent of whether the variations in energy per unit solid angle are 
caused either by the viewing angle of a standard event (provided that
$d\Gamma/d\theta_v$ is small for $d\theta_v \le \Gamma^{-1}$) or by
collimated jets, in which the typical Lorentz factor (or energy per
unit solid angle) is higher when the jet is highly collimated.

\subsection{Highly collimated outflows}

As discussed above, the gamma-rays we receive come only from material
whose motion is directed within an angle $\Gamma^{-1}$ of our line of sight.
At observer times of more than a week, the shocked outflow
starts to slow down. As $\Gamma$ drops after the deceleration shock,
the causal angle includes an increasing amount of the solid angle
toward the jet as well as towards the equator, so we observe emission
from a wider range of angles. The afterglow is thus a direct probe of
the geometry of the ejecta - if the outflow is beamed, we expect a
downturn  in the light curve as the edge of the jet becomes visible
(Rhoads 1997).  Collimation factors of $\Omega_j / 4 \pi < 0.01$ have
been derived from the observation of such steepenings\footnote{The
above argument assumes that the breaks observed in many GRB afterglow 
light-curves are due to a geometrical beam effect and
not to either a transition to non-relativistic expansion (Huang, Dai
\& Lu 2000) or an environmental effect such as a sharp density
gradient (Chevalier \& Li 2000; Ramirez-Ruiz et al. 2001). An
important feature produce by this jet collimation is the achromaticity of
the afterglow break, which will clearly distinguishes it from the
steepening that may be produced by the passage of a spectral break
through the observing band.} (Kulkarni et
al. 1999; Castro-Tirado et al. 1999; Harrison et al. 1999; Frail et
al. 2001; Panaitescu \& Kumar 2001). Although several methods of
producing and collimating jets have been proposed,  these  highly
collimated outflows of plasma with velocities close to the speed of
light  may be a unique feature of rapidly rotating,
gravitationally confined plasmas that are threaded by a strong
magnetic field (e.g. Uchida \& Shibata 1985).  
An important result from present numerical simulations of
Poynting flux-dominated outflows, which was not predicted by
semi-analytical steady state studies, is that the jets arising from
these outflows are likely to be highly collimated (see Meier, Koide \&
Uchida 2001 for a recent review). Because of the strong pinch that develops, a
narrow jet that delivers its thrust in a narrow solid angle may be a
common ingredient of strong rotating magnetic fields, not only for
accretion disks but also perhaps for many jet-producing objects. 

The
range of parameter space where high jet speeds are expected (i.e.  jets
propelled at speeds much larger than the local escape velocity) is
determined by the condition $L_{\rm MHD} \ge L_{\rm C}$. The MHD luminosity
\begin{equation}
L_{\rm MHD}={B_p^2 R_0^4 \Omega_0^2 \over 32c} 
\end{equation}
depends only on the poloidal magnetic field $B_p$ protruding from the
jet-production region at radius $R_0$ and on the angular velocity of the field
$\Omega_0$ (Blandford \& Znajek 1977), while the critical luminosity $L_{\rm
C}$ is defined to be that given by the ratio between the energy needed
for the plasma to reach the escape velocity and the free-fall time,
\begin{equation}   
L_{\rm C}= 4\pi R_0^2 \rho_0 \left({GM \over R_0}\right)^{3/2},
\end{equation}
where $\rho_0$ is the density of the plasma. The condition  $L_{\rm
MHD} \ge L_{\rm C}$ is a necessary requirement to explain the $\Gamma> 100$
outflow seen in GRBs, even from the vicinity of black holes. This
criterion is similar to that of LeBlanc and Wilson 1970, in which the
magnetic field grows to a dynamically important value in less than a
dynamical time, which seems to be a condition of forming MHD jets in
collapsing supernova cores (Meier et al. 1976;
Wheeler et al. 1999). Although carrying more power, these highly
collimated outflows will be much less efficient in imparting energy
and momentum to the outer stellar layers and may not explode (Khokhlov
et al. 1999). These failed explosions may be similar to the failed
supernova model for GRBs (MacFadyen \& Woosley 1999), continuing to
accrete much of the stellar mantle and thus collapsing into a black
hole.\\
While the very narrow $\approx 2^o$ and fast jets
$\Gamma >100$ that are speculated to exist in $\gamma$-ray bursts
(Panaitescu \& Kumar 2001) may be a general feature of strong rotating
magnetic fields, some observations may be difficult to
reconcile with this interpretation. For example, several afterglows
have shown evidence of 
large amount of X-ray line emitting material, possibly arising from ionized
iron (Piro et al. 1999; Yoshida et al. 1999; Piro et al. 2000;
Antonelli et al. 2000). The large
Fe~K$\alpha$ equivalent widths inferred from the X-ray observations
favor models in which the line is produced when the primary X-ray
emission from the source strikes Thomson-thick material and Compton
scatters into our line of sight (Lazzati, Campana \& Ghisellini 1999;
B\"ottcher \& Fryer 2000; Vietri et al. 2000; Rees \& M\'esz\'aros
2000; Ballantyne \& Ramirez-Ruiz 2001, Lazzati et al. 2001). 
These conditions impose strong
constraints on the location and geometry of the 
optically thick reflecting material, as well as on the structure of
the outflow producing the primary X-rays  - illumination of the
stellar remnant may be complicated if the GRB emission is highly
beamed. This problem may be overcome in particular source geometries
for which lower luminosities and softer spectra are expected at the
edges of the relativistic outflow (Ballantyne \& Ramirez-Ruiz
2001). More importantly, highly collimated outflows bring the GRB rate
to within a factor of 100 or so of the supernova rate (Frail et
al. 2001). Those stars that produce GRBs are likely to have core masses $>
6 M_\odot$ at the time they  collapse (Woosley 1993). Stars satisfying
this criterion are $\approx 30-50$ times rarer than those producing
\astrobj{SN Ib/c} (Izzard, Ramirez-Ruiz \& Tout 2002, in
preparation). Because 
burst formation is also likely to be favored by rapid rotation and
low metallicity (MacFadyen \& Woosley 1999), highly collimated
outflows could face some difficulties in  explaining large event rates
(i.e. $\approx 1/100$ of the supernova rate).  On the other hand,
one appealing aspect of massive star progenitors is that the large
variety of stellar parameters could explain the  diversity of jet opening
angles.  If however the \astrobj{GRB 980425}/\astrobj{1998bw}
association is real, then we 
may have a new class of GRB with lower energy $E_\gamma \approx
10^{48}(\Omega_j/4\pi)$ erg, which is only rarely observable even
though its comoving density could be substantial.

\subsection{Anisotropic relativistic outflows}

Broader jets can also be  produced by a magneto rotational mechanism,
however, some concern has been raised that these outflows may not be
easy accelerated to $\Gamma >10$ (Meier et al. 2001 and references
therein). Detailed predictions are nonetheless still in need of three
dimensional, high-resolution core-collapse calculations. Other
mechanisms of jet production such as neutrino radiation in the context
of the collapsar model (MacFadyen \& Woosley 1999) or intense
radiation of a newly born pulsar (Blackman \& Yi 1998) could be less
constraining in their beaming requirements. 

If the large variety of
behaviors exhibited from burst to burst is produced by viewing an
anisotropic universal beam configuration from different directions,
the opening angle of the jet $\theta_j > \theta_v$ should still
affect the light curve at the time when the edge of the jet becomes
visible (this break would be
additional to the one predicted from the viewing angle dependence;
Rossi et al. 2002, Zhang \& M\'esz\'aros 2002). If the critical
Lorentz factor is less than 3 or so (the opening angle exceeds 20$^o$)
such a transition might be masked by the transition from
ultra relativistic to mildly relativistic  flow, so quite generally it
would be difficult to limit the opening angle of this anisotropic
outflow if it exceeds 20$^o$. Although, the jet sideways expansion in this
model is likely to be very important and more detailed calculations are
required in order to  understand the exact shape of the afterglow
light curve, it is nonetheless important to mention that some
afterglows are unbroken power laws for over 100 days
(e.g. \astrobj{GRB 970228}),
implying that the opening angle of the late-time afterglow at long
wavelengths is probably $> 20^o$. 

Also,  under this interpretation,
the $\gamma$-rays are likely to be more narrowly beamed than the
optical afterglow and therefore there should be many ``orphan''
afterglows. The transient sky at faint magnitudes is poorly
understood, but there are two major searches to find supernova down to
R=23 (Garnavich et al. 1998; Perlmutter et al. 1998). These searches
are sensitive to afterglows of the brightness levels observed to date
and have covered a few tens of square degree years of exposure. It
then follows that the afterglow rate is not more than a few times
0.1/sq deg/yr. Considering that the magnitude limit of these searches
allows optical counterparts brighter than 1 ph cm$^{-2}$ s$^{-1}$
to be detected,  the ratio of orphan afterglows to GRB is unlikely to
exceed $\approx 10$ (M\'esz\'aros, Rees \& Wijers 1999). This condition
is not in disagreement with the few  number of orphan afterglows
that are predicted from an anisotropic universal beam configuration
(Rossi et al. 2002).  Studies of the transient sky at magnitudes down
to R $\approx$ 25 should be able to disprove or consolidate this
anisotropic model. The location of GRBs within their parental galaxy
could  help settle this question, since recently it was suggested 
that bursts located closer to the center of their parent galaxies have smaller
isotropic equivalent energies (Ramirez-Ruiz, Lazzati \& Blain
2002). If confirmed in further host observations, this correlation 
will strongly complicate this universal beam interpretation.  
A direct test of this model remains, however, the time dependent
measurements of polarization.

\section{Internal Shocks: A Phenomenological Study}

\subsection{Model Outline}
We simulate GRB light curves by adding pulses radiated in a series of
internal shocks that occur in a transient, unsteady relativistic
wind (Rees \& M\'esz\'aros 1994). 
Several authors have modeled this process by randomly selecting
the initial conditions at the central site (Kobayashi et al. 1997;
Daigne \& Mochkovitch 1998; Pilla \& Loeb 1998; Panaitescu, Spada,
M\'esz\'aros 1999, Spada, Panaitescu \& M\'esz\'aros 2000; 
Ramirez-Ruiz, Merloni \& Rees 2001). The model
used here is similar to 
that described by Ramirez-Ruiz et al. (2001) but differs in the
following aspects: (i) Our treatment of the radiation emission takes
into account synchrotron emission and Compton scattering; (ii) The
effect of photon diffusion through the optically thick wind is
included, which may be very important in obscuring pulses that occur
at small radii; (iii) We use the shock jump equations to determine
the physical conditions in the shocked fluid, and 
(iv) The duration of the pulse width
$\Delta t_0$ is calculated by adding in quadrature the angular
spreading time $\Delta T_{\rm a} \approx R/(2\Gamma^2c)$, the
radiative cooling time $\Delta T_{\rm r} \approx t_{\gamma}/\Gamma$ (where
$t_\gamma$ is the lab-frame radiative timescale), the shell crossing time
$\Delta T_{\Delta} \approx t_{\Delta} / \Gamma^2$ (where $t_\Delta \approx
\Delta/|v_{\rm sh}-v_{0}|$ is the lab-frame shock's crossing time, 
$v_{\rm sh}$ is the shell preshock flow velocity and $\Delta$ is the
shell thickness)  and the diffusion time $\Delta T_{\rm d}$ through
the optically thick wind as done by Panaitescu et al. 1999 and Spada
et al. 2000.  

The wind is discretized into a sequence of $N=T_{\rm
dur}/\;\overline{\delta t_i}$ shells with a range of
initial thickness $\Delta_i$, where $T_{\rm dur}$ is the duration time of the
wind ejection from the central source and $\overline{\delta t_i} \ll T_{\rm
dur} $ is the average interval between consecutive ejections.
The time interval $\delta t_{i}$ between two
consecutive ejections $i$ and $i+1$ is assumed to be proportional to
the $i+1$th shell energy. This implies that longer periods of
quiescence, during which the engine accumulates fuel,  are
followed by more energetic shells (this is motivated by the
observations  of Ramirez-Ruiz \& Merloni 2001 and Nakar \& Piran
2002). The different values of $\delta t_{i}$ are selected from the log normal
distribution of pulse intervals found by Norris et al. (1996).

We calculate the radii where shells collide 
and determine the emission features for each
pulse. If some inner shell moves faster than an outer one ($\Gamma_i >
\Gamma_j$), it will overtake the slow one at a radius
$R_i(t_{ij})=R_j(t_{ij})=R_c \propto \Gamma_i^2$.  
For each collision between two shells there is a reverse and a
forward shock. The  shock jump equations determine the physical parameters
of the shocked fluids: the velocities of the shock fronts,
the internal energy in the shocked fluid frame $u'$, and both the
thickness $\Delta_{ij}$ and Lorentz factor $\Gamma_{ij}$ of the 
merged shell at the end of the collision. We 
assume that in between two consecutive collisions the thickness of the
shell increases proportionally to the fractional increase of its radius
${ d\Delta_i \over \Delta_i } \propto { dR \over R }$ (e.g. Spada et
al. 2000). The ejection parameters determine the dynamics of the wind and the 
dynamical efficiency, $\epsilon_{ij}$. This efficiency reflects
the differences between the Lorentz factors of a pair of colliding
shells. The efficiency for an
individual collision can be calculated from the initial and final bulk
energies, 
\begin{equation}
\epsilon_{ij}=1- {M_{ij}\Gamma_{ij} \over M_i\Gamma_i + M_j\Gamma_j}
\end{equation}   
where
\begin{equation}
\Gamma_{ij}^2=\Gamma_i\Gamma_j{M_i\Gamma_i + M_j\Gamma_j \over
M_i\Gamma_j + M_j\Gamma_i}, 
\end{equation}
and the resulting mass is $M_{ij}= M_i + M_j$.
The first collisions remove the initial random differences between the 
Lorentz factors of successive shells. If
the mean Lorentz factor $\bar{\Gamma}$ remains steady for the entire
burst duration, then the efficiency steadily decreases during the wind
expansion. If  $\bar{\Gamma}$ is modulated
on a timescale much smaller than the overall duration of the
wind, dynamically efficient collisions at large radii are still
possible. 
 
A fraction $\zeta \le 1$ of the
shock-accelerated electrons is assumed to have a power law
distribution $-p$ in the electron Lorentz factor $\gamma_e$, starting
from a low random Lorentz factor $\gamma_m \approx { 1 \over 3}{m_p
\over m_e}{\epsilon_e \over \zeta}(\Gamma'_{ij}-1)$, where the energy stored in
electrons is a fraction $\epsilon_e$, $\Gamma'_{ij}$ is the 
Lorentz factor for internal motions in the shocked frame and we
assume $p=2.5$. The 
magnetic field is assumed to be turbulent  and parameterized through
the fraction $\epsilon_B$ of the internal energy it contains: $B^2
\approx 8 \pi \epsilon_B u'$ (primed quantities are measured
in the comoving frame). 
The synchrotron spectrum
of each pulse is approximated as  three power-law
segments, with the high energy slope(s) depending on $p$ and the
relative values of the peak and cooling frequencies. The typical
energy of synchrotron photons as well as the synchrotron cooling time
depend on the Lorentz factor $\gamma_e$ of the accelerated electrons
and on the strength of the magnetic field. The two energy release
parameters $\zeta$ and $\epsilon_B$  alter the peak energies of the
synchrotron and inverse Compton spectra (they are higher for lower
$\zeta$) and the Comptonization parameter (which is larger for smaller
$\epsilon_B$). \\
The shock-accelerated
electrons radiate, and the emitted photons can be upscattered on the
hot electrons ($\gamma_e >> 1$) or downscattered by the cold ones
($\gamma_e \approx  1$). The optical depth to upscattering, far from
the Klein-Nishina regime, is $\tau_{\rm ic} \sigma_{\rm Th} \zeta n'_e
{\rm min}(ct'_\gamma, \Delta')$ (Spada et al. 2000), where $n'_e$ is
the electron density and $t'_\gamma =t'_{\rm sy}/(1+y)$ is the
radiative timescale with $t'_{\rm sy}$ the synchrotron cooling time
and $y$ the Comptonization parameter ($y=\gamma_m^2 \tau_{\rm ic}$ for
$\tau_{\rm ic}<1$).  A fraction min(1,$\tau_{\rm ic}$) of the 
synchrotron photons is inverse Compton scattered $n_{\rm ic}$=max(1,
$\tau_{\rm ic}^2$) times unless the Klein-Nishina regime is reached
(Panaitescu  et al. 1999). 
The energy of the upscaterred photons is $h\nu_{\rm ic}={\rm
min}[\gamma_m m_ec^2, h\nu_{\rm syn}(4\gamma_m^2/3)^{n_{\rm
ic}}]$. The optical thickness $\tau_c$ for the cold electrons within
the emitting shell is calculated by taking into account the cold
electrons within the hot fluid - those that were accelerated but have
cooled radiatively while the shocked crossed the shell- and those
within the yet unshocked part of the shell.  When
$\tau_c >1$, photons are downscattered by the cold electrons before
they escape the emitting shell, leading to a decrease in photon energy
and increase in photon duration. We approximate
the increase in pulse duration due to the diffusion through optically
thick shells by the time $\Delta T_{\rm d}$ it takes for a photon to
diffuse through them (Spada et al. 2000). The addition of
all pulses gives both the burst spectra and the gamma-ray light
curve. The latter is binned on a timescale of 64 ms and is used to
computed the burst variability, that is, the average
mean-square of the count variations relative to a smoothed time
profile (calculated using a boxcar function
with a timescale equal to 30\% of the duration that contains 90\% of
the total counts; see Fenimore \& Ramirez-Ruiz 2002a). A simulated
profile depends on the specific values of the relevant model 
parameters describing the wind ejection ($N \approx 50$, $\delta t_{i}$,
$\Gamma_{\rm max}$, and  $\Gamma_{\rm min}$) and the
radiative efficiency of the pulses ($\epsilon_e$, $\zeta$
and $\epsilon_B$). 

\subsection{On the internal structure of the relativistic wind}

We study the effects of varying the the energy per unit solid angle of
the expelled outflow by an increase either in the bulk Lorentz factor
or in the baryon loading. One simple possibility is that the central engine
ejects consecutive shells in which the Lorentz factors $\Gamma_i$ are randomly
selected between $\Gamma_{\rm min}$ and  $\Gamma_{\rm
max}$. The shell mass $M$ is then assumed constant and determined by the wind
isotropic equivalent energy by requiring that
$Mc^2=E_{4\pi}/\sum^{N}_{i=1}\Gamma_i$. In this scenario, narrow jets (or
jets viewed close to the jet axis) should have higher Lorentz
factors. Assuming  $\Gamma_{\rm max} \propto \theta^{-2}_{j,v}$, then
$E_{4\pi} \propto \Gamma_{\rm max}$ (Kobayashi et al. 2002). 
Nonetheless, it could be also possible that
changes in the energy per unit solid angle 
are mainly caused by variations in the baryon loading of the
wind. In this case, the average bulk Lorentz factor
$\overline{\Gamma}=\Gamma_{\rm min}/2+\Gamma_{\rm max}/2$ remains
unchanged throughout the entire wind and equal to
$\overline{\Gamma}=E_{4\pi}/\sum^{N}_{i=1}M_ic^2$ with $M_i$ randomly
drawn from the interval $[M_{\rm min},M_{\rm max}]$ (the shell Lorentz
factors $\Gamma_i$ are random between $\overline{\Gamma}$-$\Delta
\overline{\Gamma}$ and $\overline{\Gamma}$+$\Delta
\overline{\Gamma}$). Assuming a correlation  
between $M_{\rm max}$ and $\theta_{j,v}$, then  $E_{4\pi} \propto M_{\rm
max}$. These two extreme scenarios will be referred to in the
following as {\it impulsive} and {\it lethargic } outflows,
respectively.

\subsection{Impulsive outflows}

To simulate the effects of an {\it impulsive} outflow, we calculate 
the temporal and spectral profiles of 10$^2$ realizations for a given
$\Gamma_{\rm max}$, which we used 
to evaluate the peak of the spectrum in $\nu F_\nu$ and the mean
variability. The wind (isotropic
equivalent) energy, in all cases, is drawn from a log normal
distribution with an average value
$\overline{E_{4\pi}}=10^{54}(\Gamma_{\rm max}/10^3)$ erg  and  a
dispersion $\sigma_E = 10^{1/3}$ erg. Numerous
collisions happen during the evolution of the wind. Each collision
produces a pulse, whose strength strongly depends on $\Gamma_{\rm
max}$, since the wind luminosity increases with $\Gamma_{\rm
max}$.  

The optical thickness
$\tau_c$ is determined mainly by the
wind luminosity, $L_w=E_{4\pi}/T_{\rm dur}$, and the
range of Lorentz factors in the wind.  
For small collision radii $R \propto \Gamma^2$, the 
pulse duration is dominated by the time it takes the photon to
propagate through all the shells of optical thickness above
unity. Thus, photons are downscaterred by the cold electrons before
they escape the wind, leading to an increase of the pulse width and a
decrease in the photon energy - photon diffusion through the wind widens
pulses that would otherwise would appear narrow since they occur at
small radii (Figure 2). This would be the case when the typical Lorentz factor
$\Gamma_i$ is relatively small.  The smoothing effect is expected to
be stronger when the the density of scattering is amplified by new
$e^\pm$ pairs are formed in the originally thick scattering medium
(Thompson \& Madau 2000;  Guetta, Spada \& Waxman 2001; M\'esz\'aros,
Ramirez-Ruiz \& Rees 2001; Beloborodov 2002). 
For larger radii, the pulse duration is
dominated by both shell crossing and radiative timescales (Figure 2),
which decrease as $\Gamma_i$ increases ( $\Delta T_{\Delta}$ increases
as a result of the shell widening, while  $\Delta T_{\rm r}$ is larger
for later collisions because $B$ and $\gamma_m$ are lower). 
The latter timescale is found to be comparable to 
the angular spreading one during the whole wind expansion, for the
assumed linear shell-broadening between collisions. This implies that
for large  Lorentz factors, internal collisions will produce  pulses with
small  $\Delta T_0$ and high $E_{4\pi}$ -- giving rise to a very
spiky and luminous light curve.  

By  running models with $10^2$ different values of $\Gamma_{\rm max}$
ranging from 100 to $3 \times 10^3$ (assuming  $\Gamma_{\rm
max}/\Gamma_{\rm min}$=20), we have found that  more variable profiles
($> \Gamma_{\rm max}$) have higher luminosities and can reasonably
fit the observational data (see Figure 3, where the shaded region depict the
1$\sigma$ region). The variability reaches an asymptotic
value around $V \approx 0.1$ because the light curve is binned on a
timescale of 64 ms to compare with BATSE temporal resolution. On the
other hand, for very low radii (small $\Gamma_{\rm max}$), the first
collisions remove the initial random differences between the Lorentz
factors of successive shells before reaching the radius of
transparency. As a result, the efficiency steadily decreases during the
wind expansion giving rise to very under luminous burst in which 
the variability measurement saturates due to
count statistics. In this particular example, the values of the ejecta
Lorentz factors are such (i.e. $\Gamma_{\rm
max}/\Gamma_{\rm min}$ is set to be constant) that  the average
efficiency of individual collisions $\bar{\epsilon_{ij}}$ is
independent of the intrinsic luminosity of the outflow (or
$\Gamma_{\rm max}$). Within our model more efficient collisions can
alternatively be produced by increasing $\Gamma_{\rm max}$, while
leaving $\Gamma_{\rm min }$ unchanged. The difference in the shell
Lorentz factors  (i.e. $\Gamma_{ij}$) in this
case will be much larger, giving rise to more luminous pulses than in
the previous case. A smaller range of $\Gamma_{\rm max}$ values is
thus required to produce similar changes  in the burst isotropic
luminosity (see the  banded  region in Figure 3). 
The numerical results shown in Figure 3
assume that the fraction $\epsilon_B$ is distributed uniformly in
logarithmic space between $10^{-2}$ and $10^{-1}$, $n_{\rm ISM}=1{ \rm
cm}^{-3}$ (in 
this case most collisions take place before the deceleration radius of
the wind; see Fenimore \& Ramirez-Ruiz 2002b), $\epsilon_e=0.25$ and
that $\zeta \approx 0.1$.\\  

The process by which the dissipated energy is finally radiated depends
on the energy distribution of protons and electrons in the shocked
material and on the values of the comoving density and magnetic
field. The internal shocks heat
the expanding ejecta, amplifying the preexisting magnetic field or
generating a turbulent one, and accelerate electrons, leading to
synchrotron emission and inverse Compton scattering. By varying the
injection fraction   $\zeta$ one can study the relative
intensity  of the synchrotron and inverse Compton components and
determine those values that maximize the received flux in the BATSE range. For
$\zeta > 10^{-2}$ the inverse Compton emission occurs in the Thomson
regime and carries most of the burst energy in the BATSE window, while
for $\zeta < 10^{-3}$ 
synchrotron dominates over inverse Compton scattering as the latter
takes place in the Klein-Nishina regime (Panaitescu et
al. 1999). Figure 4 shows the evolution of the synchrotron (which for
$\zeta$=1 lies mainly below the BATSE window) and inverse
Compton peaks before and after downscattering for both
thick and a thin winds. For smaller collision radii (i.e. small
$\Gamma_{\rm max}$), 
the emission takes place where the wind is optically thick, leading to
a decrease in the photon energy - diffusion through the wind
decreases the photon energy that would otherwise would appear large
since they are produced at small radii where $B$ is larger.   
For the set of parameters considered here, the Thompson limit is
usually a good approximation to treat the 
downscattering of the synchrotron photons. The inverse Compton emission,
however, peaks at large comoving frame energies, and the general cross
section has to be considered. For these energetic photons, the cross
section  strongly depends on the photon energy and varies after each
photon interaction. Figure 5 shows the shifting  of  the burst
emission towards higher energies with increasing variability, due to
the increase of $\Gamma_{\rm max}$ from 100 to $3 \times 10^3$
and assuming $\Gamma_{\rm max}/\Gamma_{\rm min}$=20 (corresponding to
the internal shock parameters of the simulations shown in Figure 3). 
This increase in energy stops once the average bulk Lorentz factor is
large enough that collisions take place   where
the wind is optically thin  (see the shaded region in Figure 5). The
typical radiation energy of each pulse also strongly depends on the
resulting Lorentz factor $\Gamma_{ij}$. Thus it is possible to produce
a larger dynamic range of spectral peak energies by increasing $\Gamma_{\rm
max}$, while leaving $\Gamma_{\rm min }$ unchanged (see banded region
in Figure 5). We found that
more variable profiles have higher spectral peaks and can reasonably
fit the observational data (Lloyd-Ronning \& Ramirez-Ruiz 2002)
provided that $\zeta >10^{-2}$. 
In  this case, the synchrotron emission lies mainly below the BATSE
window and the inverse Compton component carries
most of the energy of  burst. When synchrotron emission dominates
over inverse Compton scattering, as the latter takes place in the
Klein Nishina regime (i.e. $\zeta < \, 10^{-2}$), we find that simply varying
$\Gamma_{\rm max}$ of the shells does not produced the dynamical range
of peak energies that is observed. This is mainly 
because the typical observed synchrotron frequency, which at $R_c$ gives
\begin{equation}
h\nu_{\rm syn} \propto { \epsilon_e^2
\epsilon_B^{1/2}(\Gamma'_{ij}-1)^2 L_{w}^{1/2} \over R_c (1 +z)}, 
\end{equation}
is unlikely to increase by simply  incrementing $\Gamma$ as $R_c \propto \Gamma^2$ gives $h\nu_{\rm syn}
\propto \Gamma^{-2}$ for a constant $L_{w}$, $\epsilon_B$, $\epsilon_e$ and
$\Gamma'_{ij}$.  Assuming the correlation between the Lorentz factor
and $\theta_{j,v}$, the isotropic energy itself is related to the
source expansion velocity, $L_{w} \propto E_{4\pi} \propto \Gamma_{\rm
max}$, and  the typical observed synchrotron frequency is
then\footnote{This scaling  is valid if $\epsilon_B$, $\epsilon_e$ and
$\Gamma'_{ij}$ are constant. However, it is possible to produce a
larger dynamical range of spectral energies by increasing
$\Gamma_{\rm max}$ while leaving $\Gamma_{\rm min}$ unchanged. In this
case, $\Gamma'_{ij} \propto \Gamma$ and thus $h\nu_{\rm syn}
\propto \Gamma^{1/2}$} $\propto \Gamma^{-3/2}$. This may be overcome
by implicitly assuming that the electrons are continually accelerated
in the shocked region (see Lloyd \& Petrosian 2000) or by 
postulating that the equipartition energy acquired by  the electrons
and the magnetic field 
($\epsilon_e$ and $\epsilon_B$) are dependent on
$\Gamma$. While these are strong constraints in
the emission properties, it is worthy of notice that the
correlation between variability and 
spectral energy could be significantly affected by selection effects
and more GRBs with independent spectroscopic redshifts are need it in
order to test its validity (Lloyd-Ronning \& Ramirez-Ruiz 2002). This
relationship -if true- can help shed light on the relevant emission
mechanisms in unsteady outflows.  We would also like to emphasize that
the wind optical thickness is one of the factors to which the
variability and spectral peak energy is most sensitive. Indeed, it is
reassuring that the inclusion of this effect is of the utmost
importance to reproduce the power density spectrum of gamma-ray
bursts arising from internal shocks in unsteady outflows (Panaitescu
et al. 1999, Spada et al. 2000). 

\subsection{Lethargic outflows}
Until this point we have assumed that the changes in the energy per
unit solid angle are caused solely by an increase in the bulk Lorentz factor
($M$=constant). Nonetheless, it could be possible that changes in
$E_{4\pi}$ may be mainly caused by variations in the
baryon loading of the wind.  We have run such {\it lethargic} models 
and find that simply varying the baryon loading of the wind of the
ejected shells does not reproduce the dynamic range of variability and spectral
energies that is observed. We also study intermediate
behaviors by assuming that the shell baryon loading
and its Lorentz factor are related: $\Gamma \propto M^{\alpha}$.  We
conclude that, within such models, the burst phenomenology can be
accommodated provided  that the changes in  the energy per
unit solid angle within different beams is dominated by $\Gamma$ with
$\alpha <0.2$. This is not surprising since the timescales
that determine the pulse structure and thus the variability are highly
dependent on $R_c \propto \Gamma_i^2$. An
interesting consequence is that large variations in the energy per
unit solid angle within different beam models should be
strongly dominated by changes in $\Gamma$.

\section{Discussion}
The main conclusion we can draw from the simulations presented above
is that, in the framework of the internal shocks model, there are 
realistic assumptions in which the  variable
activity of GRB is found to be correlated with absolute peak
luminosity. We attribute this correlation to a variation of the energy per
unit solid angle caused mainly by an increase in the emitting regions'
Lorentz factors in the context of {\it impulsive} relativistic
outflows. This change of the Doppler factor could
be attributable either to a standard type of event viewed from
different orientations or by changes in the collimation of GRBs
jets. Within such models, collisions between shells with higher
Lorentz factors, which are intrinsically more luminous, produce more
variable profiles. The existence of a correlation between the
characteristic photon energy in the 
cosmological rest frame and the GRB variability, on
the other hand, imposes more constraining requirements. 
The main characteristics of the modeled bursts  with the above mentioned
features are: a high electron injection fraction -- which implies that
the inverse Compton scattering dominates over synchrotron emission --
required to increase the spectral peak energy of
the larger collision radii; and a wind optical thickness to scattering
on cold electrons above unity is required to increase the duration of
the pulses and decrease  the photon energy of the collisions that take
place at smaller radii. \\

We must also remain aware of other possibilities. For instance, we may
be wrong in supposing that the main radiation mechanism assumed to be
responsible for the burst event is synchrotron or its inverse Compton
component.  It could be that the GRB emission originates from a
fireball moving out through an extended stellar envelope, along a
funnel that is empty of matter but  pervaded by thermal radiation from
the funnel walls, leading to Compton drag (Lazzati et al. 1999;
Ghisellini et al. 2001). The fireball itself remains optically thick
until it expands beyond the stellar surface (see also Ramirez-Ruiz,
MacFadyen \& Lazzati 2002). A burst with complex
time-structure could then be produced by a series of
expanding shells.  In this scenario, faster moving shells will produce
greater peak energies, as the burst emission simply reflects the
temperature of the funnel photons  up-scattered by the square of the
bulk Lorentz factor. Furthermore, these fast shells will produce
both smaller and more luminous pulses. This is because  the observed 
variability timescale, which is related to the typical size of the region
containing the dense seed photon divided by the time compression
factor, increases with $\Gamma$ and the  Compton drag is
more efficient for faster shells. 
To this end, ejection of very rapid shells will lead to bursts
produced by the ejection of very rapid shells are both more variable
and have larger peak spectral energies. \\

Much progress has been made in understanding how afterglows can arise
from a forward shock or blast-wave moving into the external medium
ahead of the ejecta, and in deriving the generic properties of the
long wavelength afterglows that follow from this (M\'esz\'aros
2001). There still remain a number of mysteries, especially regarding
the prompt emission, in particular, the formation of the ultra relativistic
outflow, its structure and the radiation process. As we have shown,
the correlations between the characteristic photon energy, the gamma-ray burst
variability  and its luminosity, as well as the jet collimation,  can
help shed light on all three of these issues.\\
\\
\ack{
We are indebted to M. J. Rees and  D. Lazzati for very useful discussions. 
We thank D. Lazzati for highlighting the relevance of
the Compton drag scenario in reproducing the burst phenomenology.
We would also like to thank E. Fenimore, S. Kobayashi, P. Kumar,
A. Panaitescu, E. Rossi and  the referee P. M\'esz\'aros for helpful insight
regarding internal-shock and non-thermal radiation calculations. ERR
acknowledges support from CONACYT, SEP and the ORS foundation.}

{}

\clearpage

\begin{figure}
\centerline{\psfig{file=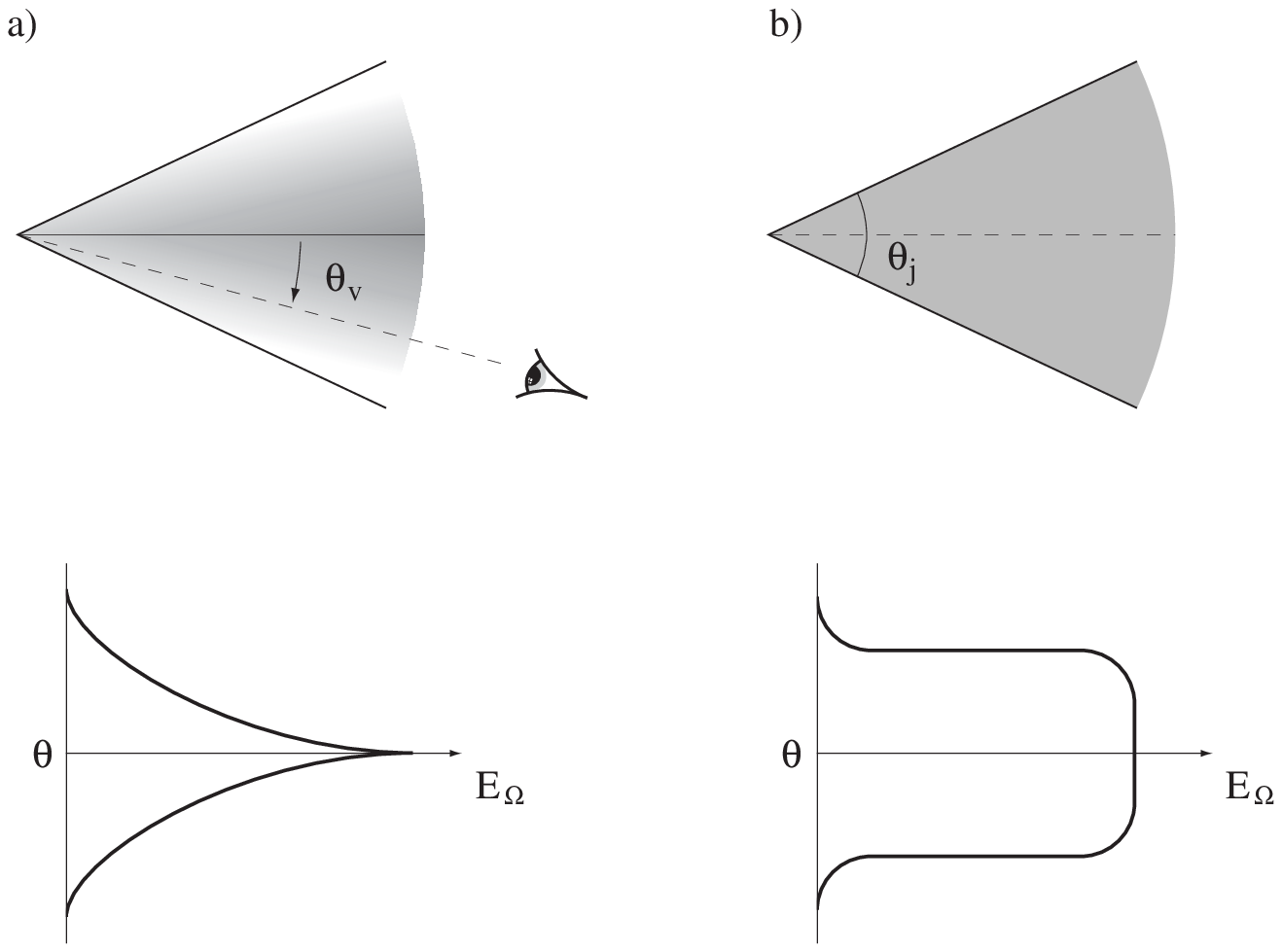,width=0.9 \textwidth}} 
\caption{{Diagram illustrating two different configurations of
beaming models. The GRB phenomenology  could
be attributable either to a standard type of event with an
axisymmetric energy distribution of the form $E(\theta_v) \propto
\theta_v^{-\beta}$ viewed from 
different orientations (a) or to a standard energy reservoir but with a
variety of beaming angles (b).} 
\label{fig1}}
\end{figure}
\clearpage

\begin{figure}
\centerline{\psfig{file=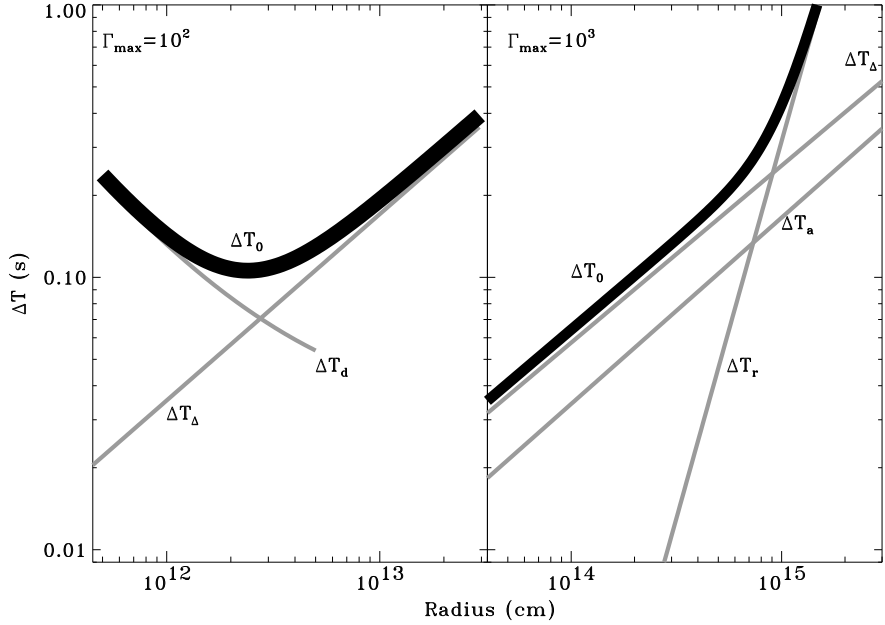,width=0.9 \textwidth}} 
\caption{{Evolution of the pulse duration as a function of the
radius of collision and of the timescales that contribute to
it. For
lower Lorentz factor $\Gamma_{\rm max} \approx 10^2$ the collisions take
place at smaller radius where the wind is optically thick, photons are
down scattered by the cold electrons before they escape, leading to an
increase of the pulse duration $\Delta T_{\rm d} \approx \Delta
T_0$ and a decrease of the pulse energy. 
For  $\Gamma_{\rm max} \approx 10^3$ must collisions occur where
the shells are optically thin $\Delta T_{\rm d} \ll \Delta
T_0$. The radiative cooling time is negligible with respect to $\Delta
T_{\rm a}$ and $\Delta T_\Delta$ for collisions occurring at small
radii, while for very large radii (i.e. small number of collisions) $\Delta
T_{\rm r}$ is the dominant contribution.} 
\label{fig2}}
\end{figure}

\clearpage

\begin{figure}
\centerline{\psfig{file=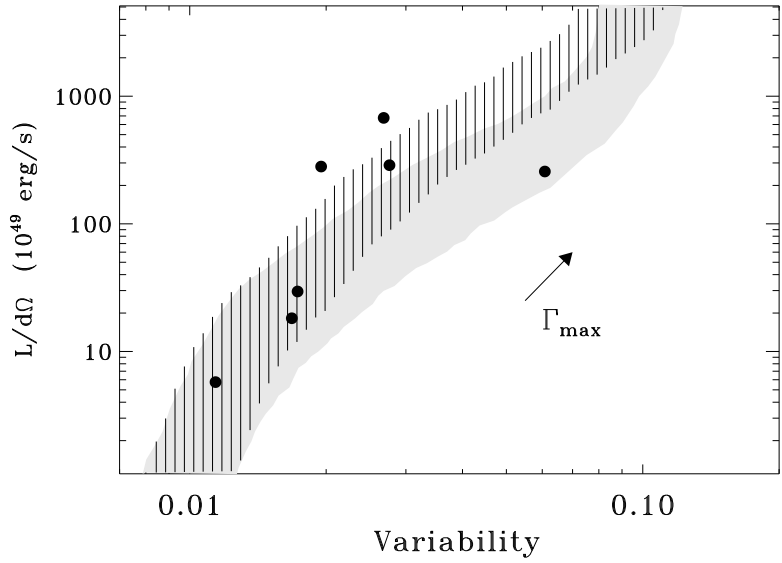,width=0.9 \textwidth}} 
\caption{{The luminosity per
unit solid angle as a function of
variability  for bursts arising from multiple shock in a relativistic
wind. The filled  circles are bursts with secure redshifts estimates
(Fenimore \& Ramirez-Ruiz 2002a). The shaded area and banded areas
represent the 1$\sigma$ regions of two  different sets of random
simulations.The former shows the effect of increasing $\Gamma_{\rm
max}/ \Gamma_{\rm min}$, while the latter depicts the effect of
increasing $\Gamma_{\rm max}$ but leaving  $\Gamma_{\rm min}$
unchanged. The observed trend reproduced as bursts with higher source
expansion velocity, which are intrinsically more luminous, produce
more variable temporal profiles.}  
\label{fig3}}
\end{figure}

\clearpage
\begin{figure}
\centerline{\psfig{file=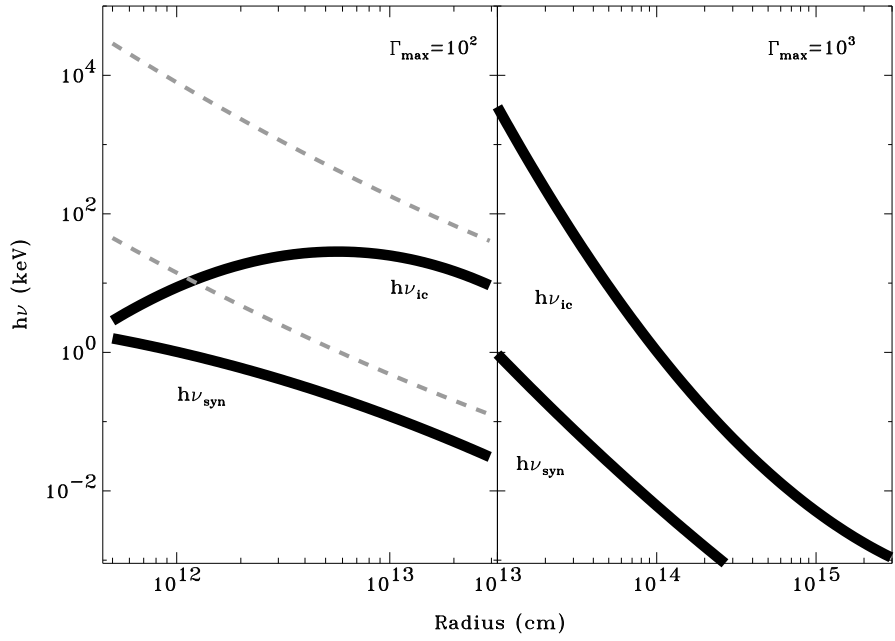,width=0.9 \textwidth}} 
\caption{{Downscattered synchrotron and inverse Compton energy
peaks as a function of the radius of collision. The dash lines
represent the evolution of the spectral peak of these two emission
components before  downscattering ($\epsilon_e=0.25$,
$\epsilon_B=0.1$, and $\zeta=1$). For collisions taking place at smaller
radii (left panel), the wind is optically thick and photons are
downscattered by the cold electrons before they escape the wind.}  
\label{fig4}}
\end{figure}

\clearpage
\begin{figure}
\centerline{\psfig{file=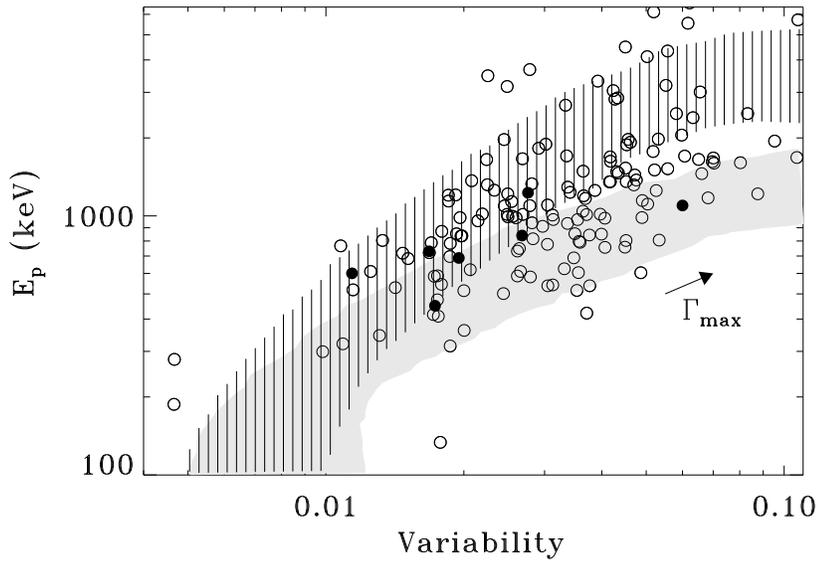,width=0.9 \textwidth}} 
\caption{{The peak of the $\nu F_\nu$ spectrum as a function of
variability  for bursts arising from an unsteady relativistic
wind. The filled 
circles are bursts with secure redshifts estimates, while the empty circles are bursts in which the  redshift is derived using the variability-luminosity
distance indicator (Lloyd-Ronning \& Ramirez-Ruiz 2002).
The shaded and banded areas represent the 1$\sigma$ regions
corresponding to the internal shock parameters of the simulations
shown in Figure 3. Again, the observed trend is reproduced,  as bursts
with higher source expansion velocity, which are 
intrinsically more luminous, radiate photons with higher
characteristic energies after diffusion through the
optical wind has taken place.}  
\label{fig5}}
\end{figure}

\end{document}